\documentclass{article} 

\title{Interpretable Enzyme Function Prediction \\ via Residue-Level Detection}
\date{}
\author{
  Zhao Yang, 
  Bing Su\thanks{\texttt{Corresponding author: subingats@gmail.com}}, 
  Jiahao Chen, 
  Ji-Rong Wen \\  
  Gaoling School of Artificial Intelligence, \\  
  Renmin University of China  
}

\usepackage{lipsum}
\usepackage[utf8]{inputenc}
\usepackage[T1]{fontenc}    
\usepackage[sf]{noto}
\usepackage[bookmarks=true, colorlinks=true,linkcolor=blue!70,citecolor=green!80!black]{hyperref}
\usepackage[numbers]{natbib}
\usepackage{wrapfig}
\usepackage{booktabs}  
\usepackage{longtable}
\usepackage{multirow}
\usepackage{ducksay}
\usepackage{etaremune}
\usepackage{afterpage}
\usepackage{capt-of}
\usepackage[table, x11names]{xcolor}
\usepackage{decorule}
\usepackage{scalerel,xparse}
\usepackage{enumitem} 
\usepackage[top=25mm, bottom=25mm, left=25mm, right=25mm,
            foot=5mm, marginparsep=0mm]{geometry}
\usepackage{svg}

\usepackage{booktabs} 
\usepackage[normalem]{ulem} 
\usepackage{graphicx}
\usepackage{subcaption}

\usepackage{float}  
\usepackage{caption}

\usepackage{graphicx}
\usepackage{tikz}
\usepackage{tikz-cd}
\usepackage{hf-tikz}
\usepackage{pgfplots} 
\pgfplotsset{compat=1.17} 
\pgfplotsset{
        table/search path={figures/drawings},
    }
\usetikzlibrary{fadings}
\usetikzlibrary{shapes, arrows, fit, backgrounds, arrows.meta}

\usetikzlibrary{matrix}
\usetikzlibrary{shadows.blur}
\usetikzlibrary{patterns, tikzmark}
\usetikzlibrary{decorations.pathreplacing, calc, decorations.markings,}
\usetikzlibrary{positioning}

\usepgfplotslibrary{groupplots}
\usepgfplotslibrary{patchplots}

\definecolor{bg}{gray}{0.97}
\definecolor{olive}{rgb}{0.6, 0.6, 0.2}
\definecolor{sand}{rgb}{0.8666666666666667, 0.8, 0.4666666666666667}
\definecolor{wine}{rgb}{0.5333333333333333, 0.13333333333333333, 0.3333333333333333}
\definecolor{deblue}{RGB}{11,132,147}
\definecolor{ocra}{RGB}{204, 119, 34}
\definecolor{emph_color}{RGB}{12, 100, 210}

\definecolor{darkersteelblue}{RGB}{25, 89, 140}

\usepackage{lettrine}
\usepackage{Typocaps}

\LettrineTextFont{\itshape}
\usepackage{amsthm}

\usepackage{minitoc}

\newcommand{\chapref}[1]{\hyperref[#1]{Chapter \ref{#1}}}
\newcommand{\secref}[1]{\hyperref[#1]{Section \ref{#1}}}
\usepackage{marginnote}

\usepackage[many]{tcolorbox}
\tcbuselibrary{breakable,xparse,skins}

\usepackage{newunicodechar}
\newunicodechar{λ}{{$\mathtt\lambda$}}
\newunicodechar{μ}{{$\mathtt\mu$}}

\usepackage{xparse}
\DeclareTColorBox{emphbox}{O{black}O{0cm}}{
    enhanced jigsaw,
    breakable,
    outer arc=0pt,
    arc=0pt,
    colback=white,
    rightrule=0pt,
    toprule=0pt,
    top=0pt,
    right=0pt,
    bottom=0pt,
    bottomrule=0pt,
    colframe=#1,
    enlarge left by=#2,
    width=\linewidth-#2,
}

\DeclareTColorBox{emphbox}{O{black}O{0cm}}{
    empty,
    breakable=true,
    outer arc=0pt,
    arc=0pt,
    rightrule=0pt,
    leftrule=2pt,
    borderline west={2pt}{0pt}{#1},
    toprule=0pt,
    top=0pt,
    right=-3pt,
    bottom=0pt,
    bottomrule=0pt,
    colframe=#1,
    enlarge left by=#2,
    width=\linewidth-#2,
}
\usepackage{listings}
\makeatletter
\newcommand\BeraMonottfamily{%
  \def\fvm@Scale{0.85}
  \fontfamily{fvm}\selectfont
}
\makeatother

\usepackage{pifont}
\usepackage{amsmath, amssymb, amsfonts, mathtools}
\usepackage{physics}
\usepackage{cancel}
\usepackage{thmtools, thm-restate}
\usepackage{accents}
\usepackage{bm}

\makeatletter
\newcommand{\ostar}{\mathbin{\mathpalette\make@circled *}}
\newcommand{\make@circled}[2]{%
  \ooalign{$\m@th#1\smallbigcirc{#1}$\cr\hidewidth$\m@th#1#2$\hidewidth\cr}%
}
\newcommand{\smallbigcirc}[1]{%
  \vcenter{\hbox{\scalebox{0.77778}{$\m@th#1\bigcirc$}}}%
}
\makeatother

\usepackage{pict2e, picture}
\makeatletter
\DeclareRobustCommand{\Arrow}[1][]{%
\check@mathfonts
\if\relax\detokenize{#1}\relax
\settowidth{\dimen@}{$\m@th\rightarrow$}%
\else
\setlength{\dimen@}{#1}%
\fi
\sbox\z@{\usefont{U}{lasy}{m}{n}\symbol{41}}%
\begin{picture}(\dimen@,\ht\z@)
\roundcap
\put(\dimexpr\dimen@-.7\wd\z@,0){\usebox\z@}
\put(0,\fontdimen22\textfont2){\line(1,0){\dimen@}}
\end{picture}%
}
\makeatother

\usepackage{scalerel}
\DeclareMathAlphabet{\nummathbb}{U}{BOONDOX-ds}{m}{n}

\makeatletter
\DeclareRobustCommand\widecheck[1]{{\mathpalette\@widecheck{#1}}}
\def\@widecheck#1#2{%
    \setbox\z@\hbox{\m@th$#1#2$}%
    \setbox\tw@\hbox{\m@th$#1%
       \widehat{%
          \vrule\@width\z@\@height\ht\z@
          \vrule\@height\z@\@width\wd\z@}$}%
    \dp\tw@-\ht\z@
    \@tempdima\ht\z@ \advance\@tempdima2\ht\tw@ \divide\@tempdima\thr@@
    \setbox\tw@\hbox{%
       \raise\@tempdima\hbox{\scalebox{1}[-1]{\lower\@tempdima\box
\tw@}}}%
    {\ooalign{\box\tw@ \cr \box\z@}}}
\makeatother

\usepackage{lineno} 

\begin{document}
\maketitle

\begin{abstract}
Predicting multiple functions labeled with Enzyme Commission (EC) numbers from the enzyme sequence is of great significance but remains a challenge due to its sparse multi-label classification nature, i.e., each enzyme is typically associated with only a few labels out of more than 6000 possible EC numbers. However, existing machine learning algorithms generally learn a fixed global representation for each enzyme to classify all functions, thereby they lack interpretability and the fine-grained information of some function-specific local residue fragments may be overwhelmed. Here we present an attention-based framework, namely ProtDETR (\textbf{Prot}ein \textbf{De}tection \textbf{Tr}ansformer), by casting enzyme function prediction as a detection problem. It uses a set of learnable functional queries to adaptatively extract different local representations from the sequence of residue-level features for predicting different EC numbers. ProtDETR
not only significantly outperforms existing deep learning-based 
enzyme function prediction methods, but also provides a new interpretable perspective on automatically detecting different local regions for identifying different functions through cross-attentions between queries and residue-level features. Code is available at \url{https://github.com/yangzhao1230/ProtDETR}.

\end{abstract}

\doparttoc

\vspace{1cm}
The development of genome sequencing technologies has unveiled a vast collection of protein sequences, but detailed functional annotations are only available for a very small number of them~\citep{acids2021uniprot}. Evaluating the functions of protein sequences via wet experiments is time-consuming, labor-intensive, and expensive, underscoring the critical need for computational methods to predict protein functions. This is particularly acute in the study of enzymes, which catalyze various biological reactions and are central to understanding metabolic processes. For the most widely-used EC number classification scheme, each class of enzyme function is assigned an EC number, which is a four-level hierarchy reflecting the intricate organization of enzyme functions. Since enzymes may have multiple functions, enzyme function prediction is essentially a multi-label classification problem. There are more than 6000 EC numbers in total and each enzyme is only associated with a few EC numbers, so it is very challenging to accurately classify or recall a sparse variable number of labels from the huge class space.

Existing conventional enzyme function prediction methods can be categorized into three groups: sequence-based~\citep{altschul215lipman, altschul1997gapped}, homology-based~\citep{steinegger2019hh}, and structure-based~\citep{zhang2017cofactor}. Since homology information and accurate structures of enzymes are difficult to obtain, it is especially appealing to predict associated EC numbers directly from enzyme sequences. Conventional sequence-based methods such as the Basic Local Alignment Search Tool for proteins (BLASTp)~\citep{altschul1997gapped, stephen1990basic}, derive function annotations solely based on sequence similarity, which can result in less reliable predictions when the sequence similarity is low.

Recent years have seen significant advances in sequence-based enzyme function prediction driven by deep learning. HDMLF~\citep{shi2023ecrecer} combines multiple sequence alignment with neural networks via a multi-task learning framework for EC number prediction. CLEAN~\citep{yu2023clean} utilizes contrastive learning to address imbalances in the training dataset, showing remarkable predictive performance on EC numbers with sparse training instances. Although the prediction performance is improved, these models lack interpretability in their prediction processes. 

To alleviate this issue, ProteInfer~\citep{sanderson2023proteinfer} introduces a degree of interpretability through convolutional neural network-based class activation mapping. EnzBert~\citep{buton2023enzbert} and DeepECtransformer~\citep{kim2023deepectransformer} leverage the Transformer Encoder to extract protein representations and attain a level of interpretability through attention scores. However, EnzBert is tailored primarily towards annotating mono-enzymes. On the other hand, DeepECtransformer employs a traditional multi-classification framework, i.e., it simplistically performs a series of binary classification tasks based on global protein representations. 

These deep learning-based methods actually follow the classification framework, i.e., they generally extract a fixed global representation for each enzyme and feed it into classifiers or compare it with templates of different EC numbers for classification. However, different enzyme functions largely depend on different local structures corresponding to specific active sites or residue fragments. If a local fragment or a weighted combination of residues in an enzyme determines a function, the fine-grained information of that fragment or combination is more effective in identifying that function. For the same enzyme, different functions may depend on different residue fragments and the discriminative fine-grained information is also different. Such fine-grained residue-level or fragment-level information may be overwhelmed in the global protein-level representation. Moreover, gaining the interpretability at the detailed granularity for in-depth analysis becomes infeasible due to the loss of residue-level features. In multi-label image classification, efforts have been made to learn different features for different classes. Query2label~\citep{liu2021query2label} introduces learnable queries, referred to as \textit{class queries}, which localize areas relevant to different objects within an image by cross-attention. 
Nevertheless, this approach is not directly transferable to multifunctional enzyme annotation, as the quadratic complexity of attention calculations for lengthy enzyme sequences and the vast number of queries corresponding to all EC numbers result in prohibitively high computational costs.

We cast multi-label enzyme function prediction as a detection problem and introduce a novel framework, ProtDETR, by leveraging the advancements in object detection, as exemplified by DETR~\citep{carion2020end}. Analogous to detecting all objects of interest in an image and determining their classes and locations, ProtDETR detects all functional residue fragments for determining relevant EC numbers. It preserves the sequence of residue-level features and employs \textit{functional queries} to identify a specific enzymatic function or denote its absence from the sequence. Unlike \textit{class queries} utilized in query2label, where more than 6000 queries are required, ProtDETR only learns 10 \textit{functional queries} as the maximum number of annotated functions for all enzymes in Uniprot is less than 10. In this way, ProtDETR adaptively extracts fine-grained fragment-level representations from residue distributions located by different queries for classifying different functions, while significantly reducing the computational demands.

With the fine-grained detection paradigm, ProtDETR achieves state-of-the-art (SOTA) results in predicting both multifunctional and monofunctional enzymes. Specifically, in multifunctional enzyme prediction tasks, ProtDETR maintains precision on par with existing SOTA methods while significantly improving the recall and F-score. For instance, on the New-392 dataset, ProtDETR achieves a recall of 0.6083, marking a 25\% improvement over CLEAN. This is pivotal for identifying potential multifunctional enzymes and uncovering the comprehensive functions of inadequately studied enzymes. ProtDETR also surpasses the specialized SOTA method, EnzBert, across all metrics and EC number hierarchies in monofunctional enzyme annotation tasks. Furthermore, cross-attentions between residue-level features and function queries in ProtDETR offer potential for EC number-specific interpretability. This not only sheds light on understanding the prediction mechanism of our model and significantly enhances the reliability and usefulness of predictions, but also provides new clues to advance the study on the fine-grained catalytic mechanism of multifunctional enzymes. These findings underscore the effectiveness of leveraging the detection framework in enzyme function prediction, signifying a methodological shift that promises to fuel future research.

\section*{Results}  
\subsection*{Framework Development and Evaluation} 

\begin{figure}[ht]
    \centering
    \includegraphics[width=1\linewidth]{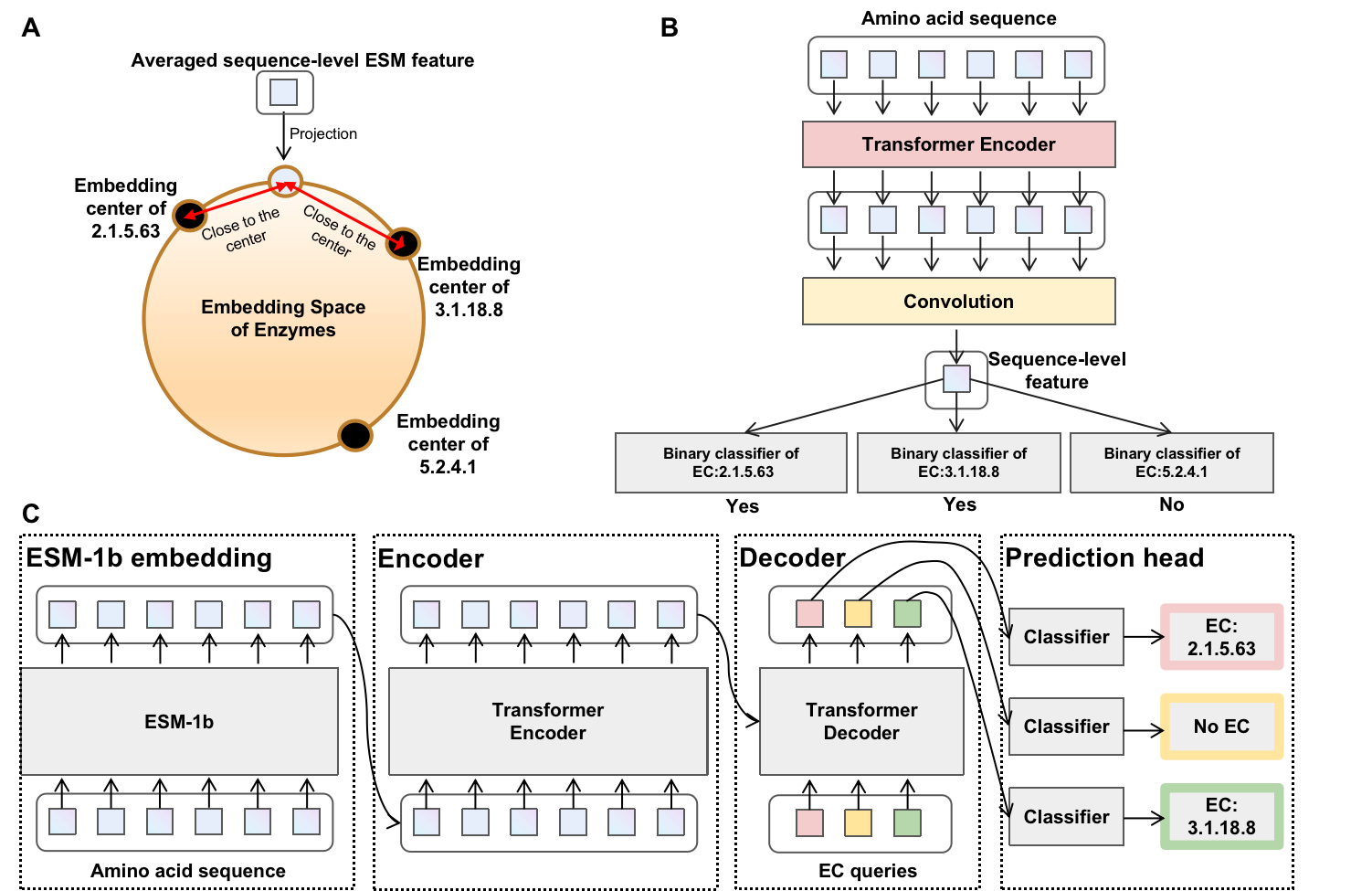}
    \captionsetup{font=small}
    \caption{\textbf{Three distinct approaches to multifunctional enzyme annotation.} (A) CLEAN utilizes contrastive learning to project enzyme features into a refined subspace, assigning EC numbers based on centroid distances. It adeptly addresses the long-tail distribution challenge of multifunctional enzymes but falls short in providing mechanistic interpretability, relying solely on global sequence-level representations. (B) DeepECtransformer leverages self-attention to model amino acid interactions, offering granularity and partial interpretability. However, in the presence of an extreme long-tail distribution, most classifiers receive insufficient training, and all functions are also predicted using the same global representations, limiting fine-grained information about specific active sites or functional fragments. (C) Our ProtDETR reconceives enzyme function prediction as a residue-level detection problem, employing unique functional queries that interact granularly with amino acids via the encoder, thus predicting EC numbers with distinct, function-specific features. It not only offers detailed, function-specific interpretability through attention score analysis but also uses each query to adaptively extract fine-grained feature interactions for precise enzyme function prediction.}
    \label{fig:model_arch}
\end{figure} 

State-of-the-art methods generally represent each enzyme as a fixed protein-level feature and then perform multi-label classification, where the protein-level feature is obtained by aggregating all residue-level features. For instance, CLEAN~\citep{yu2023clean} (Fig.~\ref{fig:model_arch}(A)) applies supervised contrastive learning to protein-level features extracted from ESM-1b~\citep{rives2021esm}, clustering enzymes according to their EC numbers in a low-dimensional subspace while differentiating between distinct EC numbers. This approach, however, lacks interpretability due to its reliance on protein-level features. DeepECtransformer~\citep{kim2023deepectransformer}, illustrated in Fig.~\ref{fig:model_arch}(B), extracts amino acid-level features with a Transformer Encoder layer. These features are then averaged to form the global protein-level representation, which is fed into multiple classifiers. The number of classifiers equals to the total number of EC numbers. This method allows for the identification of significant sites through elevated attention scores in self-attention interactions, thus providing a degree of interpretability at a coarse granularity. Conversely, EnzBert~\citep{shi2023ecrecer} adopts a similar strategy but is optimized for monofunctional enzyme classification, with the limitation of assigning only one EC number per sequence. 

We introduce a novel residue-level detection framework, namely ProtDETR, as depicted in Fig.~\ref{fig:model_arch}(C), which leverages a Transformer-based Encoder-Decoder detection architecture. For each enzyme sequence, the encoder takes embeddings extracted by ESM-1b as input and obtains the sequence of amino acid-level features. In the decoding phase, ProtDETR utilizes 10 learnable query tokens based on the assumption that enzymes seldom have more than 10 functions, and generates 10 adaptive local representations. This enables our model to effectively capture all possible EC numbers. Through cross-attention interactions between these query tokens and the amino acid sequence, ProtDETR refines the adaptive representation for each query, which mainly focuses on local residue fragments. These attended fragments may correspond to specific active or reactive sites, thereby facilitating precise classification for relevant enzyme functions. ProtDETR not only offers fine-grained and adaptive sequence modeling but also pioneers EC number-specific interpretability. It allows detailed analysis of interactions between specific queries and the amino acid sequence, revealing unique functional roles and the contribution of the amino acid composition to each enzymatic activity. This granularity in modeling and the utilization of distinct features for different predictions underscore ProtDETR's capability to illuminate the complex nature of enzymes.

During the training phase, ProtDETR conceptualizes multifunctional enzyme classification as a set prediction problem. Adapting from the DETR model~\citep{carion2020end} for object detection, ProtDETR is specially designed for enzyme function prediction, working without the necessity for bounding boxes. Adaptive query representations output by the decoder are classified to predict functions, either specifying an EC number or signaling the absence of detectable enzyme function. Our training objective aims to match these query predictions with the actual EC numbers through bipartite graph matching~\citep{kuhn1955hungarian}, establishing a direct linkage between predictions and true functions. Details of ProtDETR and its training process are presented in the Method section in the Supplementary Material.

\subsection*{ProtDETR Enables Precise Multifunctional Enzyme Classification with High Recall} 
\label{subsec:multi_func}
\begin{figure}[ht]
    \centering
    \includegraphics[width=1\linewidth]{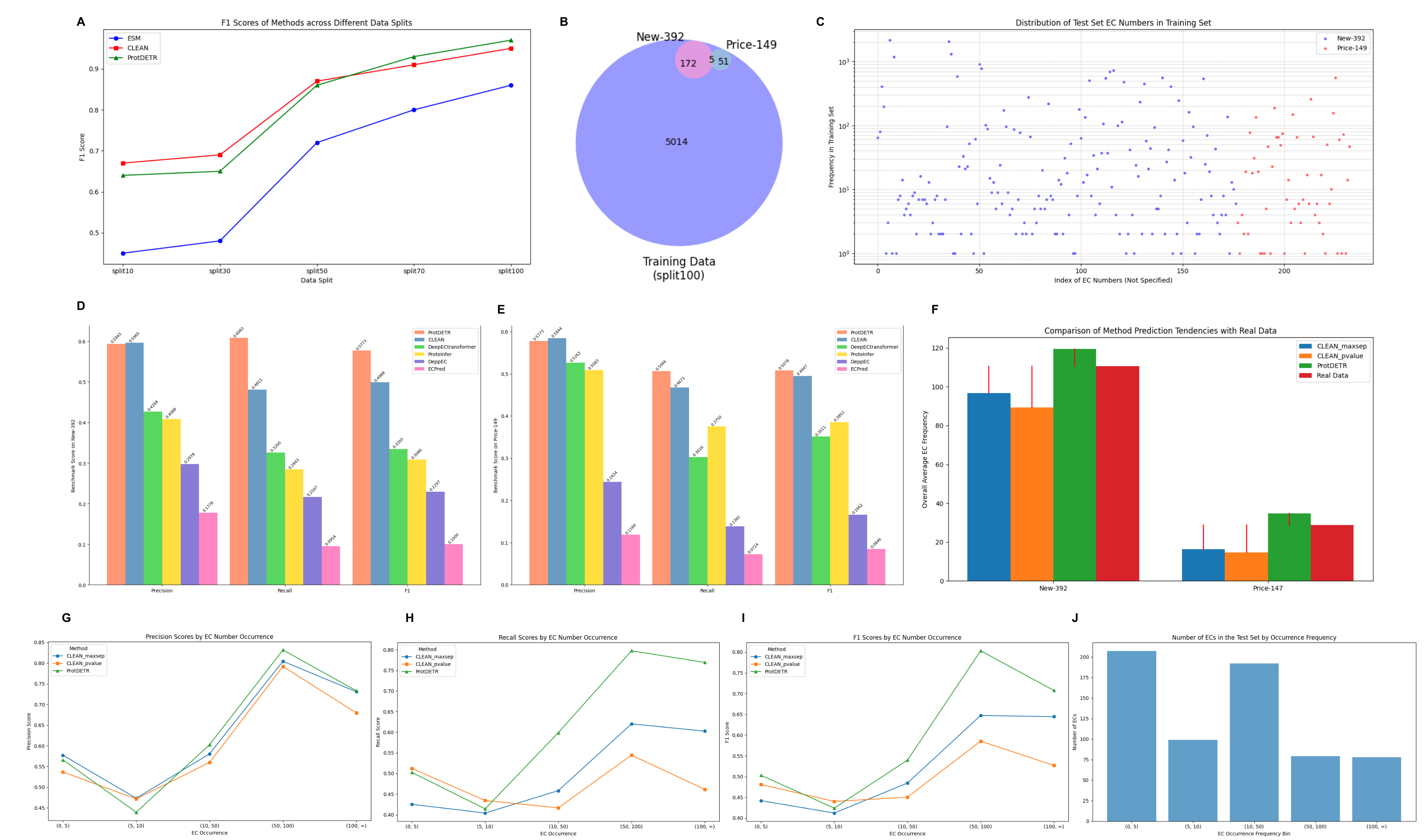}
    \captionsetup{font=small}
    \caption{\textbf{Evaluating the Multifunctional Enzyme Prediction Capabilities of ProtDETR:}
      (A) ProtDETR's performance in five-fold cross-validation shows marked improvement with larger training sets, surpassing CLEAN at higher similarity thresholds.
      (B) A Venn diagram illustrates the overlap of EC numbers between the training and test sets, highlighting shared and unique annotations.
      (C) The frequency of EC numbers in the New-392 and Price-149 test sets reveals a long-tailed distribution, with New-392 encompassing a wider array of head classes.
      (D) Performance on the New-392 dataset: ProtDETR matches the SOTA method, CLEAN, in precision and exceeds it in recall by twenty-five percent, achieving the highest performance on New-392.
      (E) Performance on the Price-149 dataset shows a consistent trend with New-392, indicating ProtDETR's robustness.
      (F) A comparison of average EC number frequencies in the training set as predicted by ProtDETR, CLEAN, and observed in the actual test sets, demonstrating ProtDETR's predictions to more closely align with the true data distribution than those by CLEAN.
      (G-I) Across various EC occurrence frequencies, ProtDETR demonstrates comparable precision to CLEAN but significantly superior recall for more frequent EC numbers, enhancing its F1 scores.
      (J) The statistical distribution of EC numbers across different occurrence ranges in the test set showcases the diversity of enzyme functionalities assessed.
    }
    \label{fig:multi_combine}
\end{figure}

To evaluate the prediction capabilities of ProtDETR for multifunctional enzymes, we adopted the experimental setting used by CLEAN \citep{yu2023clean}. Specifically, ProtDETR was trained on the split100 dataset, which is composed of about 220K instances from the expert-reviewed SwissProt section of the UniProt database. Each instance is labeled with one or more EC numbers. The model's performance was tested using two benchmarks: New-392 and Price-149. The New-392 dataset includes 392 enzyme sequences corresponding to 177 unique EC numbers, extracted from the SwissProt version released after CLEAN's training in April 2022. This dataset simulates a realistic scenario by training the model on historical SwissProt data and then predicting the functions of newly identified sequences. The Price-149 dataset, curated by ProteInfer \citep{sanderson2023proteinfer}, contains experimentally verified annotations and poses a challenge due to inconsistencies or errors in annotations by other automatic annotation methods, such as those found in KEGG \citep{kanehisa2000kegg}. 

The relationship between EC numbers in both the training and test sets is depicted in Fig~\ref{fig:multi_combine}(B), showing that each test set has a low overlap with the training set. Additionally, New-392 and Price-149 share only five common EC numbers, suggesting these datasets effectively test the model's ability to predict across a diverse range of EC numbers. Figure~\ref{fig:multi_combine}(C) illustrates the EC numbers appearing in the test sets, with the vertical axis indicating their frequency of occurrence in the training set. Each test set contains a mix of both highly frequent (head) categories, defined as EC numbers observed more than a hundred times in the training set, and infrequent (tail) categories, for example, those seen only once. Notably, only New-392 includes instances of extremely frequent categories, with occurrences exceeding a thousand times.

Before evaluating our model, we aligned with the benchmarking approach of CLEAN, using MMSeqs2 \citep{steinegger2017mmseqs2} to group data by sequence similarity, with thresholds from 10\% to 70\%. This led to the creation of datasets: split10, split30, split50, split70, and split100, where split100 represents the original, full dataset. We split each dataset with a ratio of 4:1 for training and testing, respectively, and conducted five-fold cross-validation. The average F1 score across these folds is reported. As shown in Fig~\ref{fig:multi_combine}(A), ProtDETR's F1 scores are initially lower than CLEAN's at the lower similarity levels (split10 and split30), equal at split50, and surpass CLEAN at the higher levels, with notable scores of 0.9332 versus 0.9163 for split70 and 0.9686 versus 0.9534 for split100. This improvement on larger datasets highlights the advantage of deep learning's data-driven nature, where ProtDETR's fine-grained modeling benefits from more data to enhance its prediction accuracy.

Following the hyperparameter optimization process as CLEAN, we adjusted the model's hyperparameters during the five-fold cross-validation on the split100 dataset. Once we determined a relatively suitable set of hyperparameters for training, we trained our model on the entire split100 dataset for a fixed number of epochs, in line with CLEAN's approach. We obtained ProtDETR\textsubscript{split100} for multifunctional enzyme annotation. We conducted comparisons of our model against the current SOTA methods on the test set, including CLEAN~\citep{yu2023clean}, DeepECtransformer~\citep{kim2023deepectransformer}, ProtInfer~\citep{sanderson2023proteinfer}, DeepEC~\citep{ryu2019deep}, BLASTp, DEEPre~\citep{li2018deepre}, and ECPred~\citep{dalkiran2018ecpred}. 
We evaluated our model using the same metrics as CLEAN, specifically weighted average precision, recall, and F1 score. These metrics helped assess our model's performance in multi-label classification challenges, especially in long-tail scenarios. As shown in Fig~\ref{fig:multi_combine}(D) for the New-392 dataset, our model achieved a precision of 0.5943, comparable to the SOTA performance 0.5965 set by CLEAN, while our recall of 0.6083 was 25\% higher than CLEAN's recall of 0.4811. This achievement emphasized our model's capacity for accurately annotating enzymes with potentially undiscovered functions~\citep{poirson2024proteome}. The DeepECtransformer, which also models features at the residue level like ours, performed worse in comparison to ProtDETR and CLEAN. Results for the Price-149 dataset, depicted in Fig~\ref{fig:multi_combine}(E), mirrored those for New-392, with our model matching CLEAN in precision but showing a notably higher recall of 0.5066 compared to CLEAN's 0.4671. These findings underlined the effectiveness of our approach as a tool for the annotation of multifunctional enzymes.

For a deeper analysis of the prediction outcomes, we calculated the average occurrences within the training set of EC numbers predicted by CLEAN, predicted by ProtDETR, and actually present in the test sets. As shown in Fig.~\ref{fig:multi_combine}(F)), CLEAN's predictions typically fall below the true average occurrences observed in the test sets. In contrast, while ProtDETR's predictions slightly exceed the actual averages, they align more closely. Specifically, on the New-392 dataset, CLEAN's prediction for the average frequency was 89.40, whereas ProtDETR's was closer to reality at 119.41, with the actual frequency being 110.63. For the PRICE-149 dataset, the average frequencies predicted by CLEAN and ProtDETR were 14.69 and 34.75, respectively, against an actual frequency of 28.81. 

To further explore the models' performance across various occurrence ranges, we combined the New-392 and Price-149 datasets to create a mixed dataset. We then divided the EC numbers from this dataset into several groups based on their occurrences within the split100 dataset: 0-5, 5-10, 10-50, 50-100, and over 100. Figures~\ref{fig:multi_combine}(G), (H), and (I) illustrate the precision, recall, and F1 scores for both ProtDETR and CLEAN. Our results indicated that ProtDETR's precision is comparable to that of CLEAN across all examined occurrence intervals. Notably, ProtDETR's recall significantly outperformed CLEAN's in identifying EC numbers occurring more than 10 times. This improvement was also reflected in the F1 scores, underscoring ProtDETR's enhanced ability in handling higher-occurrence, head classes without compromising performance on lower-occurrence, tail classes. Figure~\ref{fig:multi_combine}(J) provided a statistical breakdown of EC numbers across different occurrence intervals within the mixed test set. 
\subsection*{Superiority of ProtDETR Over Specialized Models in Monofunctional Enzyme Classification}  
\label{subsec:mono_func}

\begin{figure}[ht]
    \centering
    \includegraphics[width=1\linewidth]{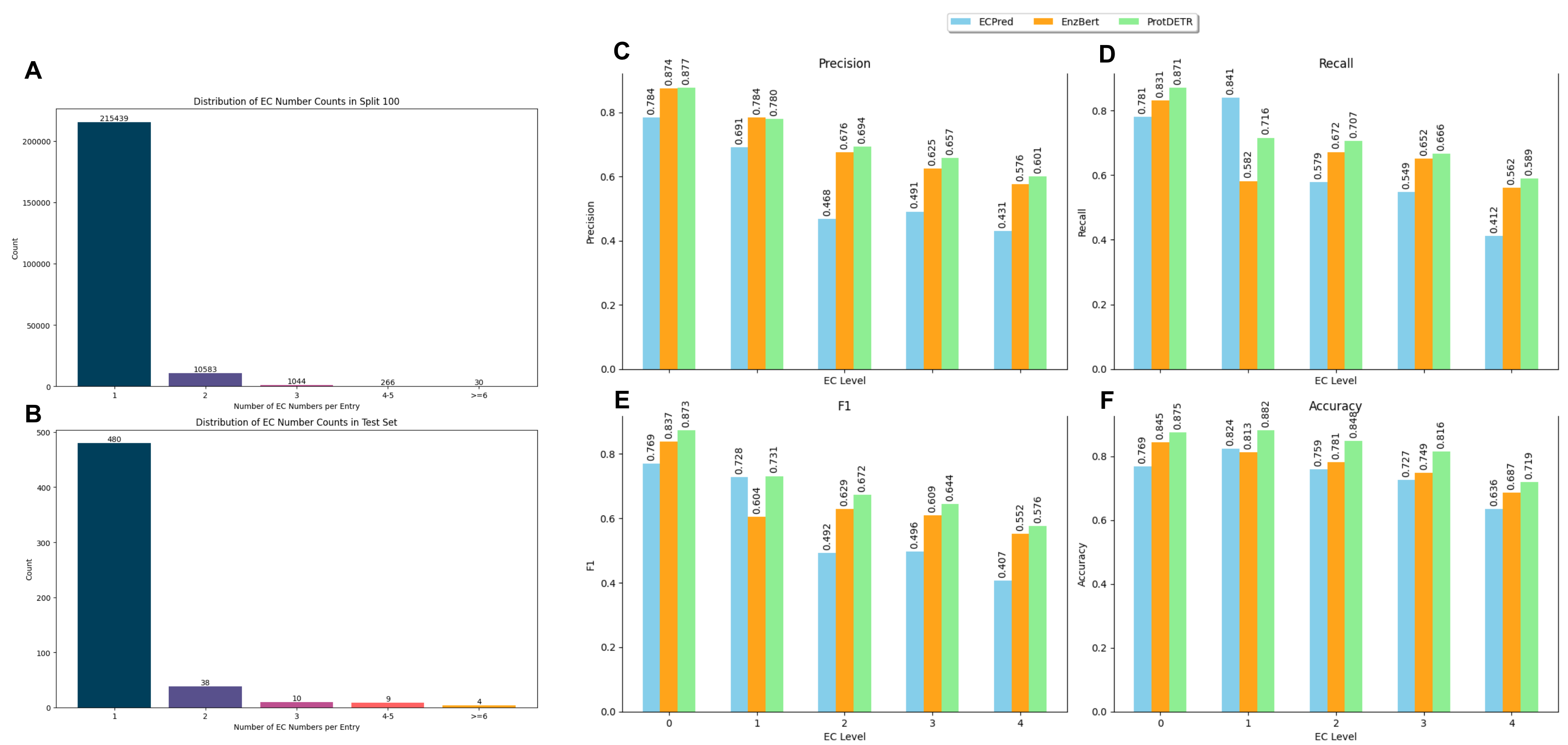}
    \captionsetup{font=small}
    \caption{\textbf{EC Number Counts Distribution in Multifunctional Enzyme Datasets and Performance on
    Monofunctional Enzymes.} (A) The distribution of EC numbers per enzyme in the split100 dataset
    indicates that a large majority of enzymes are annotated with a single EC number. (B) Statistics
    from the combined test sets of New-392 and Price-149 highlight the prevalence of monofunctional
    enzymes. (C-F) Results illustrate our method's superior performance across various EC levels on the monofunctional enzyme dataset ECPred40. Within each figure, a group represents the results on one EC level, arranged from left to right as 0, 1, 2, 3, 4, where 0 indicates the differentiation between an enzyme and a non-enzyme. ProtDETR outperforms state-of-the-art methods tailored for monofunctional enzyme classification.}
    \label{fig:mono_combine}
\end{figure}

Although enzyme function prediction was framed as a multi-label task, the majority of samples were actually single-labeled. This was demonstrated by the data in Figure~\ref{fig:mono_combine}(A), which showed that within the split100 dataset, a vast majority, 215,439 out of nearly 220,000 enzymes, were characterized by a single EC annotation. In stark contrast, only 30 enzymes exhibited more than six functionalities. This trend towards monofunctionality was also observed in the combined New-392 and Price-149 datasets, as depicted in Figure~\ref{fig:mono_combine}(B), where 480 out of 569 enzymes were identified as monofunctional. 
Given this context, the ability of a model to excel in monofunctional enzyme annotation, i.e., assigning no more than one function per protein, can significantly resolve the enzyme function prediction challenge to some extent. 

Our proposed ProtDETR was initially designed for multifunctional enzyme annotation, with the hyperparameter: the number of \textit{functional queries}, which was set to 10. To adapt it for monofunctional annotation, we simply adjusted this hyperparameter to 1, thereby transforming ProtDETR into a model specifically tailored for monofunctional annotation. We kept the chosen hyperparameters consistent with ProtDETR\textsubscript{split100}, except for adjusting the number of queries.

A detailed comparative analysis was conducted utilizing the ECPred40 dataset, which consists of a curated collection of monofunctional enzymes and some non-enzymes, assembled by EnzBert~\citep{buton2023enzbert} based on the original dataset introduced by ECPred~\citep{dalkiran2018ecpred}. Two evaluations were carried out. The first evaluation was dedicated to distinguishing between enzyme and non-enzyme classifications. In the second evaluation, we focused exclusively on the enzymes present in the testing set, assessing the accuracy of predictions across the first to fourth EC levels. These two types of evaluations are referred to as level 0 and levels 1-4, respectively. The metrics used are macro F1, precision, recall, and accuracy, following EnzBert. We trained on the training set pre-divided by EnzBert and selected the model with the best F1 score on the validation set for testing on the ECPred40 test set. The results are shown in Figure~\ref{fig:mono_combine}(C-F), where level 0 refers to distinguishing between enzymes and non-enzymes.

Remarkably, ProtDETR's performance surpassed that of EnzBert and ECPred across almost all EC levels and metrics, as illustrated in Figure~\ref{fig:mono_combine}(C-F), except for a comparable precision to EnzBert at Level 1 (0.780 vs 0.784) and a lower recall compared to ECPred at the same level (0.716 vs 0.841). Despite these specific instances, ProtDETR consistently outperformed the benchmarks, averaging an improvement of approximately 0.03 points over the nearest competitor. This signified ProtDETR's precision in not just identifying proteins as enzymes but also in its granular prediction capabilities across both broad and detailed EC categorizations. 
\subsection*{Exploring EC Number-Specific Interpretability with ProtDETR}
\label{subsec:interpret}

\begin{figure}[!htbp]
    \centering
    \includegraphics[width=1\linewidth]{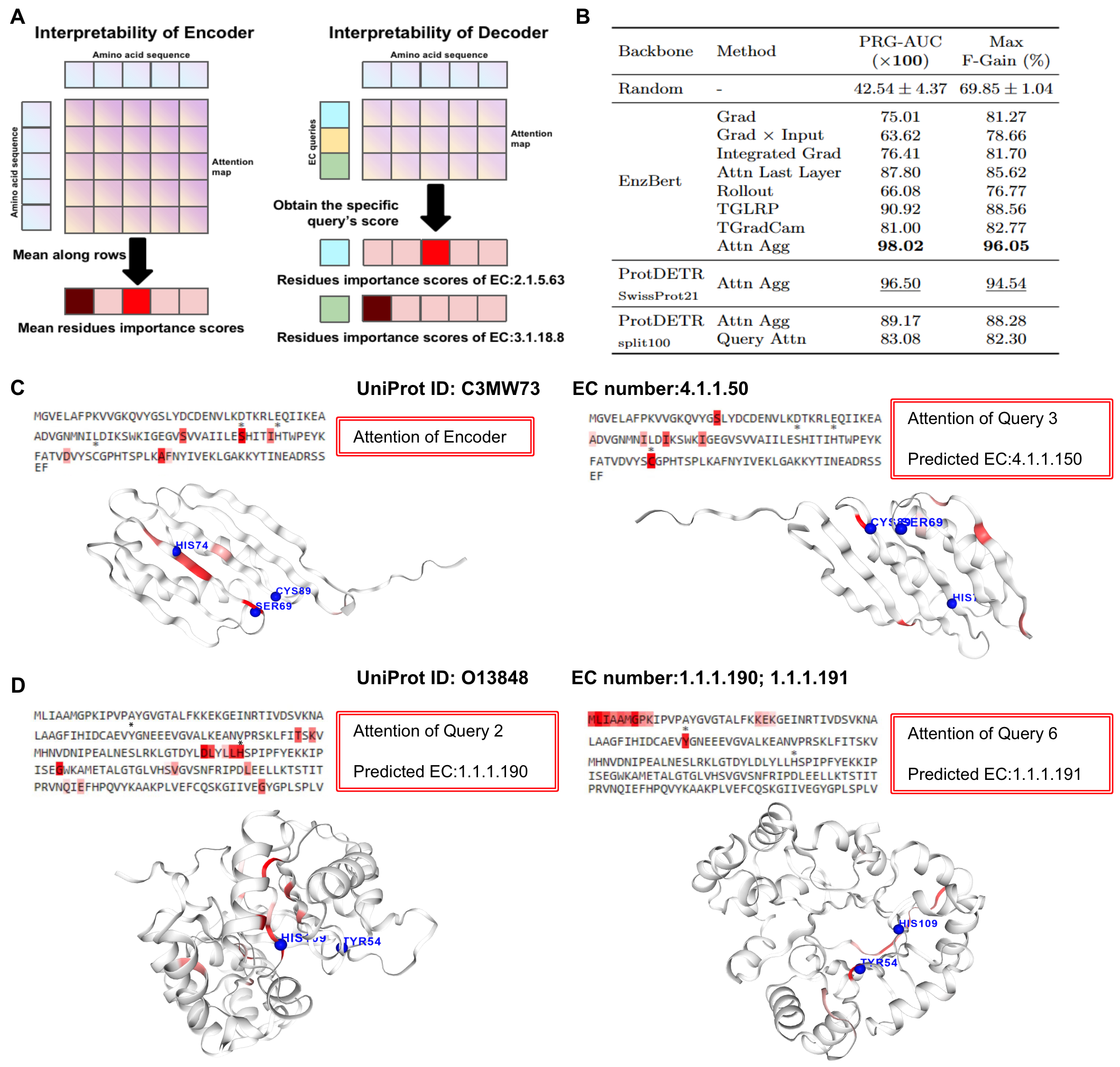}
    \captionsetup{font=small}
    \caption{\textbf{Interpretability Capabilities of ProtDETR:} (A) Illustrations on how ProtDETR's encoder and decoder identify significant sites. On the left, the encoder employs "Attn Agg" to assess the average attention each residue receives within the self-attention mechanism, highlighting general areas of interest. On the right, "Query Attn" is utilized, where the decoder employs cross-attention from each query to a residue, calculating unique focus levels of each query on different sites, providing EC number-specific interpretations. (B) Comparison of interpretability performance between ProtDETR and EnzBert on the M-CSA dataset, with \textbf{bold} indicating the best performance and \underline{underline} denoting the second best. (C) Case study for UniProt ID C3MW73 (EC 4.1.1.50). The left panel demonstrates the encoder's averaged focus while the right panel shows Query 3's targeted attention. Critical residues are highlighted in varying shades of red to signify their importance, with deeper shades indicating higher significance. Residues marked with an asterisk are identified by UniProt as active sites. (D) Case study for UniProt ID O13848 (ECs 1.1.1.190, 1.1.1.191). Both EC numbers are accurately predicted, yet they are mapped to distinct active sites. The analysis reveals unique attention patterns for queries 2 and 6, each targeting different active sites, showcasing the decoder's potential ability to interpret EC-specific functionalities.}
    \label{fig:interpret_combine}
\end{figure}

 Efforts by ProtInfer~\citep{sanderson2023proteinfer}, EnzBert~\citep{buton2023enzbert}, and DeepECtransformer~\citep{kim2023deepectransformer} to interpret enzyme function predictions hinted at deep learning's potential for discerning pivotal catalytic sites. These sites, including active and binding sites, wre thought to be the focus of heightened attention scores, suggesting areas of significant functional importance. Yet, these methods tended to provide broad, averaged insights, leaving a gap in precise, fine-grained interpretability. Figure~\ref{fig:interpret_combine}(A) introduces how ProtDETR utilized both encoder and decoder for pinpointing crucial sites. Notably, its decoder architecture paved the way for EC number-specific interpretations, moving beyond the generic insights offered by prior models. While models like EnzBert and DeepECtransformer aligned with the self-attention paradigm, ProtDETR's decoder unlocked a novel interpretative perspective by assessing cross-attention interactions. This enabled the identification of functional sites tied to particular EC numbers, underscoring ProtDETR's capability to delve into enzyme-specific functionalities.

\subsubsection*{Benchmarking on the M-CSA Database}

In a quantitative comparison, our study benchmarked ProtDETR against EnzBert, selected due to the lack of quantitative interpretability experiments from DeepECtransformer. We utilized the M-CSA (Mechanism and Catalytic Site Atlas) database~\citep{ribeiro2018mechanism}. This database documented 992 enzymes and their active sites. We applied PRG-AUC and maximum F-Gain score (Max F-Gain) metrics~\citep{flach2015prgauc}, aiming to showcase our model's precision in highlighting enzymatic functionalities and its ability to pinpoint catalytically significant residues. In simple terms, our model assigned an importance score to each residue, representing the probability of each residue being predicted as an active site. These scores were then compared with the actual labels of active sites to calculate performance metrics.

Figure~\ref{fig:interpret_combine}(B) showcases the interpretability comparison between ProtDETR and EnzBert, with a baseline established by randomizing token importance scores to evaluate effectiveness beyond random chance. EnzBert utilized a variety of interpretability methods, including neural network gradient-based techniques (such as Grad, Grad X Input, Integrated Grad~\citep{sundararajan2017intergratedgrad}, TGradCam~\citep{chefer2021tglrp}) and attention mapping approaches (like Rollout~\citep{abnar2020rollout}, TGLRP~\citep{chefer2021tglrp}). Specifically, Attn Last Layer refers to analyzing attention from the last layer of the Transformer Encoder, whereas Attn Agg involves averaging attention analysis across all layers of the Transformer Encoder. EnzBert's evaluation found that Attn Agg was the most effective method.

To ensure a fair comparison, ProtDETR\textsubscript{SwissProt21} was developed using the SwissProt21 dataset, the same dataset that EnzBert used for this task. This dataset contained approximately 500K protein sequences, including both enzymes and non-enzymes. Similarly, since this dataset comprised only monofunctional enzymes, we restricted the model to a single functional query. We adopted the Attn Agg method (shown in Figure~\ref{fig:interpret_combine}(A)) for active sites scoring. As shown in Figure~\ref{fig:interpret_combine}(B), ProtDETR\textsubscript{SwissProt21} yielded a PRG-AUC(X 100) of 96.50 and a Max F-Gain of 94.54\%, indicating that the interpretability of our model's encoder closely matched that of EnzBert. Furthermore, this significantly surpassed the performance of random guessing for active site prediction.

To investigate the interpretability capabilities of ProtDETR's decoder, we evaluated ProtDETR\textsubscript{split100} focusing on its process for classifying multifunctional enzymes. We employed a Query Attn mechanism (shown in Figure~\ref{fig:interpret_combine}(A)), which allocated importance scores based on each query's cross-attention scores to discern residue significance. However, given that the M-CSA dataset comprised monofunctional enzymes, we opted to consider only the query with the highest confidence in EC number prediction among all queries, presuming this query most likely reflected the correct prediction. Averaging attention across all queries significantly diminished performance, as most queries did not predict an EC number. Despite being trained on a dataset notably smaller than EnzBert's SwissProt21, ProtDETR\textsubscript{split100} exhibited commendable interpretability. It outperformed EnzBert's interpretative techniques in 6 out of 8 cases with the Attn Agg method and in 3 out of 8 with Query Attn. This indicated the decoder of ProtDETR also had the capability of localizing active sites.

\subsubsection*{Case Study and Visualization}

Through quantitative analysis on the M-CSA benchmark, we have shown that both the encoder and decoder of ProtDETR have considerable interpretability capabilities. However, solely relying on this quantitative data did not reveal the collaborative effects between the encoder and the various queries. Additionally, since the M-CSA dataset comprised only monofunctional enzymes, it fell short of demonstrating the interpretability capabilities for multifunctional enzymes. To address these limitations and delve deeper into the interpretability of ProtDETR in the context of multifunctional enzymes, we carried out several case studies using the UniProt database. Figures~\ref{fig:interpret_combine}(C) and (D) present two of these case studies, with more detailed information available in the supplementary materials.

In Figures~\ref{fig:interpret_combine}(C) and (D), we visually interpret the attention mechanisms for two specific enzymes. Each figure's upper section visualizes protein sequence importance, where the most critical 5\% of residues are emphasized in red, with intensity indicating significance levels. Active sites are denoted with asterisks above residues, and these sites are represented as blue spheres within the 3D structure, indicating the sequence's highlighted residues.

Figure~\ref{fig:interpret_combine}(C) illustrates ProtDETR's attention scores for Uniprot ID C3MW73, associated with EC number 4.1.1.50 and active sites at residues 69, 74, and 89. The left panel showcases the encoder's focus, particularly on active sites 64 and 79, while the right panel highlights query 3's attention, accurately predicting the enzyme's function with an emphasis on active site 89. This case underscores the collaborative identification of all pertinent active sites by both the encoder and the query that makes the accurate prediction.

Figure~\ref{fig:interpret_combine}(D) examines Uniprot ID O13848, a multifunctional enzyme linked to EC numbers 1.1.1.190 and 1.1.1.191, with active sites at 54 and 109. The left panel reveals query 2's attention, correctly inferring EC 1.1.1.190 with a focus on active site 109, whereas the right panel shows the attention of query 6, predicting EC 1.1.1.191 and concentrating on the 54. Different queries have predicted their respective correct enzymatic functions, and these two queries focus on different active sites, suggesting that ProtDETR may possess EC-number specific interpretability.

\section*{Discussion}

ProtDETR transformed the task of predicting enzyme functions into detecting these functions at the level of residues. It leverageed learnable functional queries to effectively extract and utilize local features from the sequence for predicting various EC numbers. Our extensive \textit{in silico} validation showed that ProtDETR surpassed SOTA methods (including CLEAN, DeepECtransformer, ProtInfer, BLASTp, DeepEC, DEEPre, and ECPred) in annotating multifunctional enzymes. Besides, given that most enzymes are monofunctional, our model also exceled in predicting the functions of these enzymes, outperforming the latest SOTA method, EnzBert. Additionally, ProtDETR's Encoder-Decoder architecture offered potential for EC-number-specific interpretability, identifying crucial catalytic sites, for instance, active sites. This interpretative capability was supported by both quantitative analysis and case studies.

We believe ProtDETR represents a significant step forward in the granular modeling of protein sequences, proving to be an effective tool for the challenging task of classifying multifunctional enzymes. By relying solely on sequence information for predictions, ProtDETR serves as a powerful tool for high-throughput prediction of protein functions from new sequences. Moreover, it matches the precision of the best current methods while significantly improving recall in multifunctional enzyme prediction, which is crucial for discovering enzymes' latent functions~\citep{poirson2024proteome, chen2020enzymeengineering}.

Moving beyond the constraints of previous methods, which either offered limited interpretability or none at all, ProtDETR's functional queries enable EC number-specific insights. This advancement not only deepens our understanding of the model's predictive mechanisms~\citep{christou2023time, pinney2021parallel}, enhancing the reliability and applicability of its predictions, but also supports further investigation into the roles of multifunctional enzymes~\citep{thorpe2022multifunctional, hilvert2024highmulti}.

\subsection*{Method} 
\label{sec:method}

\textbf{Problem Formulation}. Most enzymes are proteins and can be represented by amino acid sequences with length $L$. We denote such a protein sequence as $S = \{s_1, s_2, \ldots, s_L\}$, where $s_i$ is one of 20 standard amino acids. Our objective is to predict the set of active enzymatic functions for each sequence. Traditionally, this is viewed as a multi-label classification problem with a label vector $\mathbf{y} = \{y_1, \dots, y_M\}$, where $y_i = 1$ indicates activation of the $i$-th function and $M$ can be very large (often over 6000). To align with a detection-based perspective, we instead treat each protein’s functions as a \emph{set of active function indices} $Y$. We also introduce a null symbol $\varnothing$ to represent the absence of a function. Let $N$ be the maximum number of functions typically active in a single enzyme (in practice, $N=10$ is sufficient for most enzymes). When an enzyme has fewer than $N$ active functions, we pad the remaining slots with $\varnothing$. Formally,
\[
  Y = \{i \mid y_i = 1\} \;\cup\; \{\varnothing\}^{\,N - |\{i \mid y_i = 1\}|}.
\]
Hence, $|Y|=N$ is fixed, which allows us to apply a set-based detection paradigm in an end-to-end manner.

\textbf{Overview}. Figure~\ref{fig:model_arch} contrasts ProtDETR with two prior SOTA methods. CLEAN~\citep{yu2023clean} (Figure~\ref{fig:model_arch}(A)) uses global protein-level representations learned by contrastive learning~\citep{khosla2020supervisedcl}, effectively clustering enzymes by EC numbers in a latent space but lacking residue-level interpretability. DeepECtransformer~\citep{kim2023deepectransformer} (Figure~\ref{fig:model_arch}(B)) extracts amino acid features via a Transformer Encoder, then aggregates these features into a single global representation for multi-label classification. Despite partial interpretability using self-attention, it relies on thousands of binary classifiers, one for each EC class, which is challenging to train under highly imbalanced, long-tailed data.

Our ProtDETR (Figure~\ref{fig:model_arch}(C)) instead re-conceptualizes enzyme-function prediction as a \emph{residue-level detection} task. Leveraging a Transformer encoder-decoder design, we use a fixed number ($N=10$) of learnable query tokens---each query can capture the fine-grained local signatures of a possible function. Cross-attention between each query and the amino acid sequence selectively attends to crucial local fragments, such as active or binding sites, enabling more accurate and interpretable predictions. 

\textbf{ESM-1b Embedding}. We begin by encoding each protein sequence using the pretrained ESM-1b model~\citep{rives2021esm}, a large language model trained on billions of protein sequences via masked language modeling~\citep{kenton2019bert}. Given a sequence $S$, we apply ESM-1b to obtain residue-level embeddings: $F_{\mathrm{ESM}} = E(S) \in \mathbb{R}^{L \times d_{\mathrm{ESM}}}$, where $E(\cdot)$ denotes the ESM-1b embedding function. These embeddings have shown strong performance in numerous protein-related tasks~\citep{hsu2022esmif, meier2021esmmutation, hu2022evolutionaware}. Consistent with prior work~\citep{yu2023clean, shi2023ecrecer}, we truncate sequences to $L=1022$ to fit model constraints.

\textbf{Detection Transformer Architecture}. We adopt a Transformer-based encoder-decoder architecture \citep{carion2020end}, adapted from DETR for enzyme function detection. First, we project $F_{\mathrm{ESM}}$ to $d_{\mathrm{model}}$ dimensions via a linear mapping, producing feature representations that a multi-layer Transformer \emph{encoder} with $M_e$ layers refines with self-attention:
\[
  F_{\mathrm{trans}} 
  = 
  \mathrm{Linear}(F_{\mathrm{ESM}}) 
  \;\in\; 
  \mathbb{R}^{L \times d_{\mathrm{model}}}
  ,\quad
  F_{\mathrm{encoded}} 
  = 
  \mathrm{Encoder}\bigl(F_{\mathrm{trans}}\bigr).
\]
On the decoder side, we introduce $N$ learnable query tokens $Q_{\mathrm{EC}} \in \mathbb{R}^{N \times d_{\mathrm{model}}}$, each representing a potential function. A Transformer \emph{decoder} with $M_d$ layers applies cross-attention between these queries and $F_{\mathrm{encoded}}$, yielding
\[
  Q_{\mathrm{decoded}} 
  = 
  \mathrm{Decoder}\bigl(Q_{\mathrm{EC}},\,F_{\mathrm{encoded}}\bigr) 
  \;\in\; 
  \mathbb{R}^{N \times d_{\mathrm{model}}},
\]
where each query captures local context relevant to one possible enzyme function. Finally, a linear projection head maps these $N$ decoded embeddings to probability distributions over $C+1$ classes (i.e., $C$ enzyme functions plus one null class):
\[
  \mathbf{p}_i 
  = 
  \mathrm{Linear}\bigl(Q_{\mathrm{decoded}_i}\bigr)
  \;\in\;
  \mathbb{R}^{C+1},
  \quad
  \hat{y}_i 
  = 
  \mathrm{argmax}(\mathbf{p}_i),
\]
for $1 \le i \le N$, where $\hat{y}_i \in \{0,...,C\}$. 
The predicted set of enzyme functions is then defined as $\hat{Y} = \{\hat{y}_i\}$, where $\hat{y}_i = C$ indicates that the $i$-th query detects no enzyme function.

\textbf{Training Objective}. Since enzyme sequences can have multiple functions and the order of our $N$ query predictions is inherently arbitrary, we need to establish a matching between predictions and ground truth labels. Following~\citet{carion2020end}, we formulate this as a set prediction task and employ the Hungarian algorithm~\citep{kuhn1955hungarian} to find an optimal one-to-one correspondence by minimizing a global matching cost:
\[
\hat{\sigma} 
\;=\;
\arg\min_{\sigma \,\in\, \mathfrak{S}_N}
\sum_{i=1}^{N}
\mathcal{L}_{\mathrm{match}}
\bigl(Y_i,\; \hat{Y}_{\sigma(i)}\bigr),
\]
where $\mathfrak{S}_N$ is the set of all permutations over $N$ elements. Here, $Y_i \in \{1,...,C\}$ denotes the $i$-th ground truth label with $C$ representing the null class, and $\hat{Y}_{\sigma(i)}$ is the $\sigma(i)$-th predicted label. The cost $\mathcal{L}_{\mathrm{match}}$ is based on the negative log-likelihood for each ground-truth class.

Once we obtain the optimal assignment $\hat{\sigma}$, we calculate the final cross-entropy loss over matched pairs:
\[
 \mathcal{L}(Y,\,\hat{Y})
 \;=\;
 \sum_{i=1}^N
 -\log\,
 \hat{p}_{\,\hat{\sigma}(i)}\bigl(Y_i\bigr),
\]
where $\hat{p}_{\,\hat{\sigma}(i)}\bigl(Y_i\bigr)$ is the predicted probability of the correct class for $Y_i$ under the optimal matching. This ensures that each ground-truth function (including null) is paired with exactly one model prediction, enforcing a one-to-one correspondence in the multi-function setting. To address the extreme long-tail problem in enzyme function prediction, we further discuss the additional loss we used in Appendix~\ref{appendix::long_tail}.

\textbf{Localization of Key Enzymatic Sites via Attention}.  Residue-level attention scores in ProtDETR provide interpretability. First, in the final layer of the \emph{encoder}, we aggregate the self-attention scores across all heads to find residues of high global importance:
\[
 \text{AttnAgg}(i) \;=\; \sum_{h=1}^{H}\sum_{j=1}^{L}\text{EncAttn}_{h}(i,j),
\]
where $\text{EncAttn}_{h}(i,j)$ denotes the self-attention weight from residue $i$ to residue $j$ under head $h$ in the last encoder layer. 

More uniquely, ProtDETR also provides \emph{function-specific} insight via the \emph{decoder} cross-attention. For each query $q_{\mathrm{EC}}$ and residue $i$, we similarly aggregate cross-attention weights from the final decoder layer:
\[
 \text{QueryAttn}(q_{\mathrm{EC}}, i) 
 \;=\;
 \sum_{h=1}^{H} \text{DecAttn}_{h}(q_{\mathrm{EC}}, i),
\]
where $\text{DecAttn}_{h}(q_{\mathrm{EC}}, i)$ represents the cross-attention weight from query $q_{\mathrm{EC}}$ to residue $i$ under head $h$. This reveals where each \emph{individual function} query focuses in the sequence 
. As multifunctional enzymes typically have distinct sites for different activities, these query-specific attention maps provide deeper insight into the roles of particular residues in enabling diverse enzyme functions. 

\section*{Data availability}
All datasets related to the multifunctional enzyme classification, including split10, split30, split50, split70, split100, New-392, and Price-149, are available at \url{https://github.com/tttianhao/CLEAN}. The dataset for monofunctional enzyme classification, ECPred40, can be found at \url{https://doi.org/10.5281/zenodo.7253910}. The dataset for validating interpretability, M-CSA, is accessible at \url{https://www.ebi.ac.uk/thornton-srv/m-csa/}. Data used in the case study are derived from UniProt, available at \url{https://www.uniprot.org/}.

\bibliography{main.bib}

\newpage
\appendix
\section{Addressing the Long-Tailed Challenge}
\label{appendix::long_tail}

\begin{figure}[ht]
    \centering
    \includegraphics[width=1\linewidth]{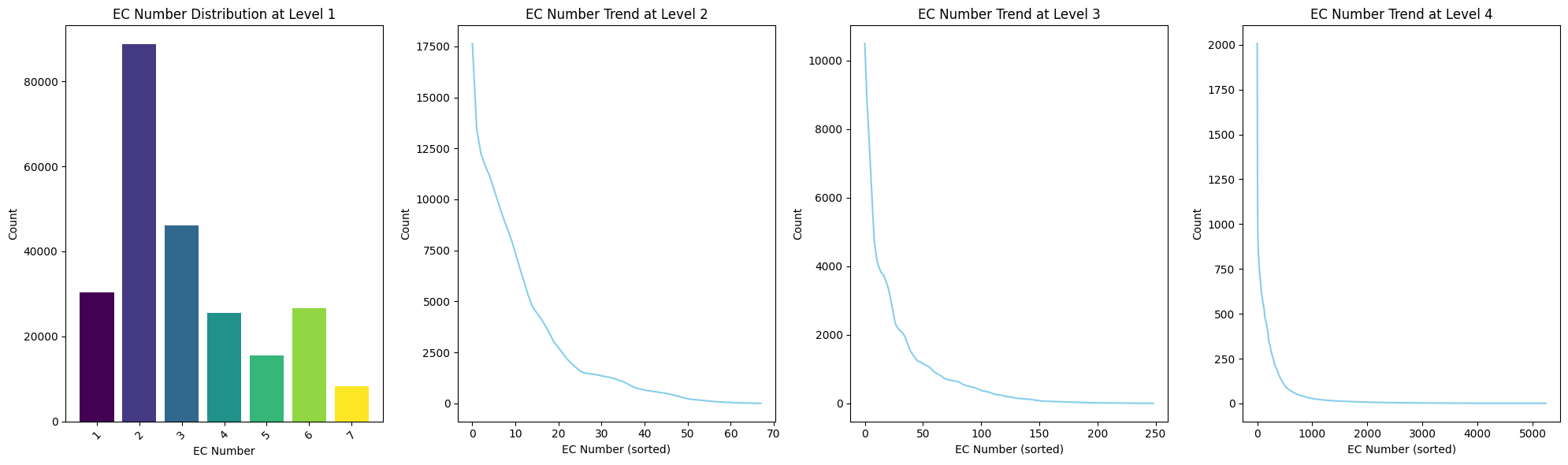}
    \caption{\textbf{Long-tailed distribution in the split100 dataset, observable across all four hierarchical levels of EC numbers.} This widespread pattern underscores the significant challenge of class imbalance in enzyme function prediction.}
    \label{fig:ec_trend}
\end{figure}

The long-tailed distribution of enzyme classes in training data presents a significant challenge in multifunctional enzyme annotation, as depicted in Figure ~\ref{fig:ec_trend}. Each EC number level exhibits a pronounced long-tailed distribution. Additionally, non-enzymatic instances significantly outnumber specific enzyme classes, exacerbating the class imbalance. In machine learning, such long-tail distributions typically result in significant performance degradation, as models tend to favor the more frequent classes at the expense of rare classes. To address this issue, we explore three distinct strategies to mitigate the effects of class imbalance and enhance the model's performance across all classes.

\textbf{Baseline Adjustment (BA):} Following DETR, we apply a constant weight to non-enzyme classes within our loss function to mitigate their predominance due to high occurrence rates. However, this method does not directly address the imbalance within enzyme classes. The weight \( w \) is a hyperparameter, set to 0.1 based on DETR guidelines.
\begin{equation}
\mathcal{L}_{\text{BA}} = -\sum_{i=1}^{N} \left( w \cdot \mathbb{1}_{\{Y_i = \varnothing\}} + \mathbb{1}_{\{Y_i \neq \varnothing\}} \right) \cdot \log(\hat{p}_i)
\end{equation}

\textbf{Inverse Frequency Reweighting (IFR):} This method corrects class imbalance by adjusting weights inversely proportional to class frequencies, thus enhancing the influence of rarer classes. The weight for each class \( w_{c} \) is set inversely proportional to its frequency in the training dataset, enhancing the model's attention to rarer classes.
\begin{equation}
\mathcal{L}_{\text{IFR}} = -\sum_{i=1}^{N} \frac{1}{\text{freq}(c_i)} \cdot \log(\hat{p}_i)
\end{equation}

\textbf{Logit Adjustment (LA):}
Logit Adjustment is a straightforward yet effective technique commonly employed in long-tail image classification tasks to enhance model performance across diversely distributed classes~\cite{menon2020adjustlogits}. By adding the logarithm of class frequencies to the logits, this method increases the difficulty of learning for high-frequency classes, aiming to balance the training across different class distributions. This logit modification forces the model to pay more attention to less frequent classes, potentially reducing the dominance of frequent classes in the loss function:
\begin{equation}
\text{adjusted logits} = \text{logits} + \log(\text{freq}(c_i))
\end{equation}
After adjusting the logits, the standard cross-entropy loss is applied.

We conducted experiments using the naive cross-entropy method (CE). We evaluated these methods on a 4:1 cross-validation using the split70 dataset, with performance based on the F1 score, as shown in Table~\ref{tab:longtail_f1}. The CE and BA methods proved to be the least effective, with F1 scores of only 0.8903 and 0.8901, respectively. This is likely because these methods did not account for the long-tailed distribution of different classes, leading to poor results. Conversely, the IFR approach outperformed the others with an F1 score of 0.9332 and was thus adopted. The third methodology, LA, while effective in long-tailed image classification tasks, was less effective in our context, yielding an F1 score of 0.9242. We speculate that datasets tailored for long-tailed image classification, often engineered with standardized declining power-law frequencies, might be ill-suited for the multifunctional enzyme classification task. This highlights the need for novel long-tail handling strategies, a promising avenue for future research.

\begin{table}[ht]
\centering
\caption{Performance comparison of the proposed methods for solving the long-tailed distribution based on F1 score.}
\label{tab:longtail_f1}
\begin{tabular}{@{}l*{1}{>{\centering\arraybackslash}p{2cm}}@{}}
\toprule
Method & F1 Score \\
\midrule
CE & 0.8903 \\
BA & 0.8901 \\
IFR & \textbf{0.9332} \\
LA & 0.9242 \\
\bottomrule
\end{tabular}
\end{table}
\section{Hyperparameter Selection Strategy}
\label{appendix::implementation_details}

The hyperparameter configurations for our model were primarily influenced by the guidelines proposed in DETR~\cite{carion2020end}, tailored to meet the specific demands of our enzyme function prediction task. DETR originally suggested using 6 encoder layers, 6 decoder layers, and 8 attention heads, with a inference threshold set at 0.7. Given the flexibility of adjusting thresholds during the testing phase, our initial focus was on fine-tuning the encoder and decoder layers as well as the number of attention heads. We evaluated three different configurations: (3, 3, 4) and (6, 6, 8) -- the standard DETR setup.

Our assessments demonstrated that models with fewer parameters generally outperformed their more complex counterparts, likely due to the prolonged training epochs required for larger models to achieve convergence. Comparative results on split100's 5-fold cross-validation for the configurations (3, 3, 4) and (6, 6, 8) are presented in Table~\ref{tab:config_336} and Table~\ref{tab:config_668}, respectively. We conducted a total of one hundred training epochs on our model. It is evident that the configuration (3, 3, 6) consistently outperformed (6, 6, 8) across all epochs and inference thresholds, also requiring fewer epochs to converge while using fewer parameters. Consequently, we have adopted the (3, 3, 4) configuration for our final hyperparameter settings in the task of multifunctional enzyme annotation.

\begin{table}[ht]
\centering
\caption{F1 scores for configuration (3,3,6) across different inference thresholds and epochs}
\label{tab:config_336}
\begin{tabular}{@{}ccccc@{}}
\toprule
Infer Threshold & 20 Epochs & 50 Epochs & 100 Epochs \\ \midrule
0.5             & 0.9486    & 0.9543    & 0.9571     \\
0.7             & 0.9551    & 0.9577    & 0.9606     \\
0.9             & 0.9617    & 0.9623    & 0.9645     \\
0.99            & 0.9600    & 0.9678    & 0.9686     \\ \bottomrule
\end{tabular}
\end{table}

\begin{table}[ht]
\centering
\caption{F1 scores for configuration (6,6,8) across different inference thresholds and epochs}
\label{tab:config_668}
\begin{tabular}{@{}ccccc@{}}
\toprule
Infer Threshold & 20 Epochs & 50 Epochs & 100 Epochs \\ \midrule
0.5             & 0.9331    & 0.9225    & 0.9629     \\
0.7             & 0.9296    & 0.9198    & 0.9576     \\
0.9             & 0.9296    & 0.9197    & 0.9505     \\
0.99            & 0.9296    & 0.9197    & 0.9387     \\ \bottomrule
\end{tabular}
\end{table}

\begin{table}[htbp]
    \centering
    \caption{Configurations of Model HyperParameters for Different Training Sets}
    \label{tab:model_params}
    \begin{tabular}{@{}lcc@{}}
    \toprule
    & ProtDETR\textsubscript{split100} & ProtDETR\textsubscript{ECPred40} \\ \midrule
    Encoder Layer & 3 & 3 \\
    Decoder Layer & 3 & 3 \\
    Num heads & 4 & 4 \\
    Number of Queries & 10 & 1 \\
    Hidden dim & 256 & 256 \\
    FFN DIM & 2048 & 2048 \\
    Dropout & 0.1 & 0.1 \\
    Learning Rate & 1e-4 & 1e-4 \\
    \bottomrule
    \end{tabular}
\end{table}

This training culminated in the deployment of our ProtDETR\textsubscript{split100} on the complete split100 dataset. We adopted a fixed epoch strategy similar to that used in CLEAN~\cite{yu2023clean}, which trained on the full split100 dataset for 7000 epochs. However, we limited our training to 50 epochs dedicated to multifunctional prediction. This decision was based on the observation that, by the 50th epoch, our model's performance in five-fold cross-validation on the split100 dataset had already significantly surpassed that of CLEAN, with no substantial gains from additional training.

A pivotal adjustment was made to the Number of Queries parameter, predicated on our analytical insight. Given DETR's framework, where an image is unlikely to host more than 70 objects, thus setting 100 queries, we reasoned that an enzyme with up to 7 or 8 functions warrants 10 queries by analogy. This foundational setting was preserved in ProtDETR\textsubscript{split100}. For ProtDETR\textsubscript{ECPred40}, tailored to monofunctional enzyme prediction, we adjusted the query count to one, reflecting the task's monofunctional focus. Due to the smaller dataset size of ECPred40, we trained the model for a total of 20 epochs. During the validation phase of ProtDETR\textsubscript{ECPred40}, peak performance was observed at the 17th epoch, which guided our decision to finalize the model at this stage.

Table~\ref{tab:model_params} delineates the specific hyperparameters employed for each model version, detailing our tailored approach to enzyme function prediction across different training sets.

\doparttoc

\end{document}